\documentclass[11pt]{article}

\usepackage{citesort}
\usepackage{graphicx}
\usepackage{amssymb}

\makeatletter
\@addtoreset{equation}{section} 

\makeatother

\renewcommand{\r}{\mathbf{r}}
\newcommand{\dd}{\mathrm{d}}
\newcommand{\e}{\mathrm{e}}
\newcommand{\ii}{\mathrm{i}}
\newcommand{\tbf}{\mathbf{t}}
\newcommand{\dbar}{\not\hspace{-0.5 mm}\partial}
\newcommand{\Tr}{\mathrm{Tr}}

\begin{document}

\title{{\bf Charge Fluctuations for a Coulomb Fluid in a Disk on a
Pseudosphere}} 

\author{{\bf B. Jancovici}$^1$ {\bf and} {\bf G. T\'ellez}$^2$} 

\maketitle

\begin{abstract}
The classical (i.e. non-quantum) equilibrium statistical mechanics of a
Coulomb fluid living on a pseudosphere (an infinite surface of constant
negative curvature) is considered. The Coulomb fluid occupies a large
disk communicating with a reservoir (grand-canonical ensemble). The
total charge $Q$ on the disk fluctuates. In a macroscopic description,
the charge correlations near the boundary circle can be described as
correlations of a surface charge density $\sigma$. In a macroscopic
approach, the variance of $Q$ and the correlation function of $\sigma$
are computed; they are universal. These macroscopic results are shown to
be valid for two solvable microscopic models, in the limit when the
microscopic thickness of the surface charge density goes to zero.
\end{abstract}

\medskip


\medskip

\noindent LPT Orsay 03-10

\vfill
 
\noindent $^1$Laboratoire de Physique Th\'eorique, B\^atiment 210, 
Universit\'e de Paris-Sud, 91405 Orsay, France (Unit\'e Mixte de 
Recherche no.8627-CNRS); e-mail: Bernard.Jancovici@th.u-psud.fr

\noindent $^2$Grupo de F\'{\i}sica T\'eorica de la Materia Condensada, 
Departamento de F\'{\i}sica, Universidad de Los Andes, A.A.4976, Bogot\'a,
Colombia; e-mail: gtellez@uniandes.edu.co

\newpage

\section{INTRODUCTION}
This paper is dedicated to Elliott Lieb on the occasion of his 70th
birthday. It is a variation on a theme to which Elliott has brought
major contributions.\cite{LL,LN}

How statistical mechanics is affected by the curvature of space might be
of some interest in general relativity, and also is an amusing problem
per se. A simple case is a two-dimensional system living on a
pseudosphere, i.e. a surface of constant negative curvature. Unlike the
sphere, the pseudosphere has an infinite area, and therefore, on a
pseudosphere, one can consider the thermodynamic limit of some system
while keeping a given curvature. A special feature is that, for a large
domain, the neighborhood of the boundary has an area of the same order
of magnitude as the whole area of this domain; this feature makes the
approach to the thermodynamic limit rather different to what happens in
a flat space.

More specifically, the present paper deals with a two-dimensional
classical (i.e. non-quantum) Coulomb fluid living on a pseudosphere. 
This is a system of charged particles interacting by
Coulomb's law, with this law defined on the pseudosphere, i.e. as the
solution of the Poisson equation written with the pseudosphere
metric. The Coulomb fluid is assumed to be in equilibrium and confined
in a large disk drawn on the pseudosphere. The grand-canonical ensemble
is used: the fluid can freely exchange particles with a reservoir. Thus,
the total charge $Q$ may fluctuate. Furthermore, there are charge
correlations which, near the circle boundary of the disk, can be
described as correlations of a surface charge density $\sigma$. The aims
of the present paper are to compute the variance of $Q$ and the
two-point correlation function of $\sigma$. It will be shown that these
quantities are universal (i.e. independent of the microscopic nature of
the fluid). These universal results will be checked on two exactly
solvable models: the two-component plasma, made of two species of
particles of opposite signs, and the one-component plasma, made of one
species of particles in a neutralizing background.

These problems have already been studied and solved in a flat space. For
a finite two-dimensional Coulomb fluid in a plane, the total charge $Q$
essentially does not fluctuate~\cite{chargefluctuations}. Furthermore,
in the case of a large disk of radius $R$ centered at the origin, the
surface charge correlation function~\cite{correlations} is given
by the universal expression
\begin{equation}
\beta\langle\sigma(\varphi)\sigma(0)\rangle^{\mathrm{T}} = -\frac{1}
{2\pi^2[2R\sin(\varphi/2)]^2} \label{plane}
\end{equation}
where $\beta$ is the inverse temperature, $\sigma(\varphi)$ the
surface charge density on the boundary circle at the point of polar
angle $\varphi$, and $<\ldots>^{\mathrm T}$ is a truncated statistical
average. These results have been checked on exactly solvable
models.~\cite{chargefluctuations,surface} 

Here, the same problems are considered, now on a pseudosphere. In
Section 2, some basic properties of the pseudosphere and of Coulomb's
law on it are recalled. In Section 3, macroscopic electrostatics on a
pseudosphere is used for determining the variance of the total charge
$Q$ and the correlation function of the surface charge density $\sigma$.
In Section 4, the results are checked on a solvable model, the
two-component plasma at a special temperature. In Section 5, the results
are checked again on another solvable model, the one-component plasma at
a special temperature.

\section{PSEUDOSPHERE AND COULOMB'S LAW}
Let us recall a few properties of the surface of constant negative
curvature called a pseudosphere. Such a surface is a two-dimensional
manifold, the entirety of which cannot be embedded in three-dimensional
Euclidean space. Its properties are defined by its metric. Several
sets of coordinates are commonly used.
 
The one which renders explicit the resemblance with the sphere is 
$(\tau,\varphi)$ with $\tau\in [0,\infty[$ and 
$\varphi\in\, ]-\pi,\pi]$, the metric being 
\begin{equation}
\dd s^2=a^2(\dd\tau^2+\sinh^2\tau\:\dd\varphi^2) \label{metric1}
\end{equation}
where $-1/a^2$ is the Gaussian curvature (instead of $1/R^2$ for a
sphere of radius $R$). The geodesic distance $s$ between two points at 
$(\tau,\varphi)$ and $(\tau',\varphi')$ is given by 
\begin{equation}
\cosh(s/a)=\cosh\tau\cosh\tau'-\sinh\tau\sinh\tau'\cos(\varphi-\varphi')
\label{distance}
\end{equation}
In particular, the geodesic distance of the point $(\tau,\varphi)$ to
the origin is $a\tau$. The Laplace-Beltrami operator is
\begin{equation}
\Delta=\frac{1}{a^2}\left(\frac{1}{\sinh\tau}\frac{\partial}{\partial\tau}
\sinh\tau\frac{\partial}{\partial\tau}+
\frac{1}{\sinh^2\tau}\frac{\partial^2}{\partial\varphi^2}\right)
\label{Laplacian}
\end{equation}
The set of points at a geodesic distance from the origin less than or 
equal to $R=a\tau_0$ will be called a disk of radius $R$. Its boundary
will be called a circle of radius $R$. Its circumference is
\begin{equation}
\mathcal{C}=2\pi a\sinh\tau_0  \label{circumference}
\end{equation}
and its area is
\begin{equation}
\mathcal{A}=4\pi a^2\sinh^2(\tau_0/2) \label{area}
\end{equation}
It is remarkable that, for a large radius, both the circumference and
the area are proportional to $\exp\tau_0$: the neighborhood of the
boundary circle has an area of the same order of magnitude as the whole
area!

Another often used set of coordinates is $(r,\varphi)$ with 
$r/(2a)=\tanh(\tau/2)$. Then, the metric is
\begin{equation}
\dd s^2=
\frac{\dd r^2+r^2\dd\varphi^2}{\left(1-\frac{r^2}{4a^2}\right)^2}
\label{metric2}
\end{equation}
When these coordinates are used, the whole (infinite) pseudosphere maps
on a disk of radius $2a$, the Poincar\'e disk.

Finally, here it will be convenient to use also the coordinates
$(\epsilon,\varphi)$ with 
\begin{equation}
\tanh(\tau/2)=\e^{-\epsilon} \label{epsilon}
\end{equation}
Then the metric is
\begin{equation}
\dd s^2=\frac{a^2}{\sinh^2\epsilon}(\dd\epsilon^2+\dd\varphi^2)
\label{metric3}
\end{equation}
and the Laplace-Beltrami operator has the simple form 
\begin{equation}
\Delta=\frac{\sinh^2\epsilon}{a^2}
\left(\frac{\partial^2}{\partial\epsilon^2}+
\frac{\partial^2}{\partial\varphi^2}\right) \label{Laplacian1}
\end{equation}

The Coulomb potential $v(s)$ at a geodesic distance $s$ from a unit
point charge obeys the Poisson equation
\begin{equation}
\Delta v(s)=-2\pi \delta^{(2)}(s) \label{Poisson}
\end{equation}
where $\delta^{(2)}$ is the Dirac distribution on the pseudosphere. The
solution of (\ref{Poisson}) which vanishes at infinity is
\begin{equation}
v(s)=-\ln\tanh\frac{s}{2a}   \label{Coulomb}
\end{equation}

\section{MACROSCOPIC ELECTROSTATICS, CHARGE FLUCTUATIONS, SURFACE
CHARGE CORRELATIONS}
\subsection{Two Problems in Macroscopic Electrostatics}
Here are two problems, the solution of which will be needed in the
following. On the pseudosphere, an ideal conductor fills the disk of
radius $R=a\tau_0$ centered at the origin.  

\textbf{Capacitance.} The first problem, a very simple one, is: What is 
the capacitance of this disk? If the disk carries a charge $Q$, this
charge uniformly spreads on its circumference and, by Newton's theorem
(which can be easily shown to be valid on a pseudosphere), generates on
the whole disk the constant electric potential
$Q\,v(R)=-Q\ln\tanh(\tau_0/2)$. Therefore, the capacitance is
\begin{equation}
C=-\frac{1}{\ln\tanh\frac{\tau_0}{2}}  \label{capacitance}
\end{equation}
In the large-disk limit $\tau_0\rightarrow\infty$,
$C\sim\exp(\tau_0)/2$.

\textbf{A Point Charge in the Presence of the Disk.} The second problem
is: A unit point charge is located, outside the disk, at point
$(\tau',\varphi'=0)$. The disk is grounded (i.e. kept at zero
potential). What is the electric potential $\phi(\tau,\varphi;\tau')$ at 
some point $(\tau,\varphi)$, outside the disk? The method of images,
which can be used for a flat disk, does not seem to work on a
pseudosphere, and a Fourier expansion will be used. 

The potential due to the unit point charge alone is (\ref{Coulomb}). 
Expressing this potential in terms of $\cosh(s/a)$, using
(\ref{distance}), and expanding as a Fourier series in $\varphi$ (in the
case $\tau<\tau'$ which suffices for our purpose) gives
\begin{equation}
v(s)=\epsilon'+\sum_{\ell=1}^{\infty}\frac
{2\sinh\epsilon'\ell\;\e^{-\epsilon\ell}}{\ell}\cos\ell\varphi \;\;\;
(\tau_0<\tau<\tau') \label{alone}
\end{equation}
where we have gone from the variables $\tau$ and $\tau'$ to the variables
$\epsilon$ and $\epsilon'$ defined by (\ref{epsilon}) and its analog for
$\epsilon'$. Using (\ref{Laplacian1}), one easily checks that the
Laplacian of each term of (\ref{alone}) vanishes.
  
The full potential in the presence of the disk is obtained by adding to 
(\ref{alone}) terms of zero Laplacian symmetrical in $\epsilon$ and
$\epsilon'$: a term of the form $A_0\epsilon'\epsilon$, and terms of the
form $A_{\ell}\sinh\epsilon'\ell\:\sinh\epsilon\ell\:\cos\ell\varphi$. 
These terms do vanish when $\tau$ or $\tau'$ goes to infinity. The
coefficients $A_{\ell}$ are determined by the condition that the
potential vanishes on the disk, i.e. when $\tau=\tau_0$. The result is
\begin{equation}
\phi(\tau,\varphi;\tau')=\epsilon'-\frac{\epsilon\epsilon'}{\epsilon_0}
+\sum_{\ell=1}^{\infty}2\sinh\epsilon'\ell\left(\e^{-\epsilon\ell}
-\frac{\e^{-\epsilon_0\ell}}{\sinh\epsilon_0\ell}\sinh\epsilon\ell\right)
\frac{\cos\ell\varphi}{\ell} \label{potential}
\end{equation}
where $\epsilon_0$ is related to $\tau_0$ by the analog of
(\ref{epsilon}).

\subsection{Charge Fluctuations}
The disk of radius $R=a\tau_0$ centered at the origin is filled with a
Coulomb fluid. It can freely exchange charges with a reservoir located
at infinity (grand canonical ensemble). If macroscopic electrostatics is
applicable, what is the variance of the charge $Q$ carried by the disk? 

The same reasoning as in the flat space case~\cite{chargefluctuations},
using linear response theory, gives for the variance
\begin{equation}
\beta\langle Q^2 \rangle^{\mathrm{T}}=C  \label{chargevariance}
\end{equation} 
where here the capacitance $C$ is given by (\ref{capacitance}). This
result (\ref{chargevariance}) just says that the variation of the energy
$Q^2/2C$ has the usual thermal average $(1/2)\beta^{-1}$. In the
large-disk limit $\tau_0\rightarrow\infty$,
\begin{equation}
\beta\langle Q^2 \rangle^{\mathrm{T}}\sim\frac{\e^{\tau_0}}{2}\sim
\frac{\mathcal{C}}{2\pi a} \label{chargevariance1}
\end{equation} 

\subsection{Surface Charge Correlations}
In three dimensions, macroscopic electrostatics deals with volume charge
densities and surface charge densities. In the present case of a 
two-dimensional system (a disk), the analog of the surface charge
density actually is a charge per unit length on the boundary circle; we
shall nevertheless still call it a surface charge density 
$\sigma(\varphi)$. If macroscopic electrostatics is applicable, what is
the two-point correlation function of $\sigma(\varphi)$?

The same reasoning as in
the case of a flat space~\cite{correlations}, using linear response
theory, gives for the two-point correlation function
\begin{equation}
\beta\langle\sigma(\varphi)\sigma(0)\rangle^{\mathrm{T}}=
-\frac{1}{(2\pi a)^2}
\left(\frac{\partial^2\phi(\tau,\varphi;\tau')}{\partial\tau\partial\tau'}
\right)_{\tau=\tau'=\tau_0}    \label{bigradient}
\end{equation}
where $\phi(\tau,\varphi;\tau')$ is the electric potential
(\ref{potential}).
 
The second derivative in (\ref{bigradient}) can be expressed in closed
form in terms of the Jacobi theta~\cite{theta} function $\theta_1$.
For the sake of dealing only with convergent series, from
(\ref{potential}) one first computes the second derivative for 
$\tau'>\tau_0$, i.e. $\epsilon'<\epsilon_0$:
\begin{equation}
\left(\frac{\partial^2\phi(\tau,\varphi;\tau')}{\partial\tau\partial\tau'}
\right)_{\tau=\tau_0}=-2\sinh\epsilon_0\:\sinh\epsilon'
\left(\frac{1}{2\epsilon_0}+\sum_{\ell=1}^{\infty}\frac{\cosh\epsilon'\ell}
{\sinh\epsilon_0\ell}\:\ell\cos\ell\varphi\right)  \label{bigradient1}
\end{equation}
For taking the limit of (\ref{bigradient1}) when $\tau'\rightarrow\tau_0$,
one substracts from and adds to $\cosh\epsilon'\ell/\sinh\epsilon_0\ell$
a term $\e^{-(\epsilon_0-\epsilon')\ell}$. The limit of one of the two
resulting series is computed after the summation has been performed:
\begin{equation}
\lim_{\epsilon'\rightarrow\epsilon_0}\sum_{\ell=1}^{\infty} 
\e^{-(\epsilon_0-\epsilon')\ell}\ell\cos\ell\varphi=
-\frac{1}{4\sin^2\frac{\varphi}{2}} \label{trivial}
\end{equation}
The other series involves $(\cosh\epsilon'\ell/\sinh\epsilon_0\ell)-
\e^{-(\epsilon_0-\epsilon')\ell}$. It remains absolutely convergent when
the limit $\epsilon'=\epsilon_0$ is taken in each term. Thus
\begin{equation}
\left(\frac{\partial^2\phi(\tau,\varphi;\tau')}{\partial\tau\partial\tau'}
\right)_{\tau=\tau'=\tau_0}=-2\sinh^2\epsilon_0\left(\frac{1}{2\epsilon_0}
-\frac{1}{4\sin^2\frac{\varphi}{2}}+2\sum_{\ell=1}^{\infty}\frac
{\e^{-2\epsilon_0\ell}}{1-\e^{-2\epsilon_0\ell}}\:\ell\cos\ell\varphi\right)
\label{bigradient2}
\end{equation}
The sum in (\ref{bigradient2}) can be expressed in terms of the Jacobi 
$\theta_1$ function, since \cite{theta}
\begin{equation}
\frac{\theta'_1(v,q)}{\theta_1(v,q)}=\pi\cot\pi v+4\pi
\sum_{\ell=1}^{\infty}\frac{q^{2\ell}}{1-q^{2\ell}}\:\sin2\ell\pi v
\label{sum}
\end{equation}
Setting $v=\varphi/(2\pi)$ and $q=\e^{-\epsilon_0}$ in (\ref{sum}), and
using its derivative with respect to $\varphi$ in (\ref{bigradient2})
gives for the correlation function (\ref{bigradient}) the closed form
\begin{equation}
\beta\langle\sigma(\varphi)\sigma(0)\rangle^{\mathrm{T}}=
\frac{1}{(2\pi a)^2}
\sinh^2\epsilon_0\left[\frac{1}{\epsilon_0}+\frac{1}{\pi}\frac{\dd}
{\dd\varphi}\frac{\theta'_1(\frac{\varphi}{2\pi},\e^{-\epsilon_0})}
{\theta_1(\frac{\varphi}{2\pi},\e^{-\epsilon_0})}\right]
\label{correlation}
\end{equation}

In the flat-space limit $a\rightarrow\infty$, $\tau_0\rightarrow 0$, 
for a fixed value of $R=a\tau_0$, it can be checked that (\ref{plane})
is recovered. More interestingly, in the opposite limit of a radius $R$ 
large compared to the ``curvature radius'' $a$, i.e. when
$\tau_0\rightarrow\infty$ and $\epsilon_0\rightarrow 0$,
(\ref{correlation}) takes a simpler form. Indeed, after a Jacobi
imaginary transformation \cite {theta,transtheta}, the $\theta_1$ 
function can be expressed as the series
\begin{equation}
\theta_1\left(\frac{\varphi}{2\pi},\e^{-\epsilon_0}\right)=
\left(\frac{\pi}{\epsilon_0}\right)^{1/2}\sum_{n=-\infty}^{n=\infty}
(-1)^n\exp\left[-\frac{\pi^2}{\epsilon_0}\left(\frac{\varphi}{2\pi}-
\frac{1}{2}+n\right)^2\right]   \label{theta1}
\end{equation}
If $0<|\varphi|<\pi$, in the small-$\epsilon_0$ limit
($\epsilon_0\sim 2\e^{-\tau_0}$), the first two
leading terms of the series (\ref{theta1}) are $n=0$ and $n=1$. When
only these terms are kept, to first order in their ratio
$\exp(-\pi|\varphi|/\epsilon_0)$, (\ref{correlation}) becomes  
\begin{equation}
\beta\langle\sigma(\varphi)\sigma(0)\rangle^{\mathrm{T}}
\sim -\frac{1}{2a^2}
\exp\left(-\frac{\e^{\tau_0}\pi|\varphi|}{2}\right)\;\;\;(0<|\varphi|<\pi)
\label{correlation1}
\end{equation}

\section{TWO-COMPONENT PLASMA ON A PSEUDOSPHERE}
The total charge fluctuation and the surface charge correlation
have been obtained under the assumption that macroscopic electrostatics 
is valid. In the large-disk limit $R\gg a$, these results 
(\ref{chargevariance1}) and (\ref{correlation1}) will now be checked on
two solvable microscopic models. Such checks are welcome, because a
two-dimensional case when macroscopic electrostatics is not valid,
unexpectedly at first sight, is known: the charge fluctuations in a
short-circuited circular condenser \cite{chargefluctuations}.

Macroscopic electrostatics uses the concept of surface charge density.
Actually, in a microscopic model, this ``surface density'' will have
some microscopic thickness, and for macroscopic electrostatics to be
valid, it is necessary that this thickness be negligible compared to the
macroscopic lengths. The microscopic model which will be used in the
present section is the
two-component plasma at a special temperature. Its microscopic scale is
characterized by a fugacity $\zeta$, with the dimension (length)$^{-2}$.
In a disk of radius $R$ on a pseudosphere with the ``radius of
curvature'' $a$, there are two dimensionless parameters involving
$\zeta$: $\zeta a^2$ and $\zeta R^2$. Necessary conditions for
macroscopic electrostatics to be valid is that both these parameters be
large compared to 1. Here, for simplicity, the disk is assumed to be
large $(R\gg a)$.

\subsection{Review of the General Formalism}
The two-component plasma is a system of two species of particles, of
charges $\pm 1$. At the special inverse temperature $\beta=2$, the model
is exactly solvable in different geometries, in particular on a
pseudosphere \cite{beta2}. For the sake of completeness, the method of
solution is briefly revisited, in a form simpler than in the original
papers, by a generalization of what has been done in the case of a
one-component plasma \cite{pressures}. 

In terms of the coordinates $(r,\varphi)$ (the Poincar\'e disk
representation), the Coulomb interaction (\ref{Coulomb}) between two
unit point charges at $\r_i$ and $\r_j$ is 
\begin{equation}
v(s)=-\ln\left|\frac{(z_i-z_j)/(2a)}{1-\frac{z_i\bar{z}_j}{4a^2}}\right|
\label{Coulomb1}
\end{equation}
where $z_j$ is the complex coordinate of particle $j$ (and $\bar{z}_j$
its complex conjugate): $z_j=r_j\e^{\ii\varphi_j}$. The interaction
(\ref{Coulomb1}) happens to be the Coulomb interaction in a flat disk of
radius $2a$ with ideal conductor walls at zero potential. Therefore, one
can use the techniques which have been developed
\cite{Forrester,JancoTellez} for dealing with ideal conductor walls.
 
When $\beta=2$, the Boltzmann factor for $N_+$ positive particles with
vector coordinates $\r_i^+$ and corresponding complex coordinates 
$z_i^+,\,1\leq i\leq N_+$, and $N_-$ negative particles with vector 
coordinates $\r_i^-$ and corresponding complex coordinates
$z_i^-,\,1\leq i\leq N_-$, can be written as (with, for the time being,
$2a$ taken as the unit of length: $2a=1$)
\begin{eqnarray}
&B_{N_+,N_-}=& \nonumber \\
&\frac{\prod\limits_{1\leq i<j\leq N_+}(z_i^+-z_j^+)
(\bar{z}_i^+-\bar{z}_j^+)\prod\limits_{1\leq k<l\leq N_-}
(z_k^--z_l^-)(\bar{z}_k^--\bar{z}_l^-)
\prod\limits_{m=1}^{N_+}\prod\limits_{n=1}^{N_-}(1-z_m^+\bar{z}_n^-)
(1-\bar{z}_m^+z_n^-)}
{\prod\limits_{1\leq i<j\leq N_+}(1-z_i^+\bar{z}_j^+)
(1-\bar{z}_i^+z_j^+)\prod\limits_{1\leq k<l\leq N_-}
(1-z_k^-\bar{z}_l^-)(1-\bar{z}_k^-z_l^-)
\prod\limits_{m=1}^{N_+}\prod\limits_{n=1}^{N_-}(z_m^+-z_n^-)
(\bar{z}_m^+-\bar{z}_n^-)}& \label{Boltzmann}
\end{eqnarray}
(in the cases $N_+=0$ and $N_-=0$, the corresponding products in
(\ref{Boltzmann}) should be replaced by 1; in particular $B(0,0)=1$). It
is convenient to define 
\begin{equation} 
B'_{N_+,N_-}=\frac{B_{N_+,N_-}}{\prod\limits_{m=1}^{N_+}
\prod\limits_{n=1}^{N_-}
(1-z_m^+\bar{z}_m^+)(1-z_n^-\bar{z}_n^-)}  
\label{Boltzmann1}
\end{equation}
$B'$ is the Boltzmann factor in a disk with ideal conductor walls at
zero potential, including now in its denominator the contribution from
the interaction of each particle with its own image. $B'$ has the
advantage that it can be written as a $N\times N$ determinant
($N=N_++N_-$ is the total number of particles), by using the Cauchy
identity
\begin{equation}
\frac{\prod\limits_{1\leq i<j\leq N}(u_i-u_j)(v_i-v_j)}
{\prod\limits_{i=1}^N\prod\limits_{j=1}^N(u_i-v_j)}=
(-1)^{N(N-1)/2}\det\left(\frac{1}{u_i-v_j}\right)_{i,j=1,\ldots,N}
\label{Cauchy}
\end{equation}
Indeed, choosing 
\begin{eqnarray}
u_i=z_i^+,   &v_i=1/\bar{z}_i^+,   &1\leq i\leq N_+       \nonumber   \\
u_{i+N_+}=1/\bar{z}_i^-   &v_{i+N_+}=z_i^-,   &1\leq i\leq N_-  
\label{labels}
\end{eqnarray}
in (\ref{Cauchy}) gives, after some simple manipulations and the
reestablishment of an arbitrary value of $2a$, a $N\times N$ determinant
\begin{equation}
B'_{N_+,N_-}=\det\,A_{ij} \label{Boltzmann2}
\end{equation}
where
\begin{eqnarray}
A_{ij}&=&\frac{4a^2}{4a^2-z_i^+\bar{z}_j^+}\;\;
\mathrm{if}\;1\leq i,j\leq N_+  \nonumber \\
A_{ij}&=&\frac{2a}{z_i^+-z_{j-N_+}^-}\;\;
\mathrm{if}\;1\leq i\leq N_+,\;\;\;N_+<j\leq N  \nonumber \\
A_{ij}&=&\frac{2a}{\bar{z}_{i-N_+}^--\bar{z}_j^+}\;\;
\mathrm{if}\;N_+<i\leq N,\;\;\;1\leq j\leq N_+  \nonumber \\
A_{ij}&=&\frac{4a^2}{4a^2-\bar{z}_{i-N_+}^-z_{j-N_+}^-}\;\;
\mathrm{if}\;N_+<i,j\leq N \label{elements}
\end{eqnarray}
If the (perhaps different) fugacities are $\zeta_+$ and $\zeta_-$ for
the positive and negative particles, respectively, the grand partition
function can be written as 
\begin{equation}
\Xi=\sum_{N_+=0}^{\infty}\sum_{N_-=0}^{\infty}\frac{1}{N_+!N_-!}\int
\prod\limits_{m=0}^{N_+}\dd^2\r_m^+\zeta_+(r_m^+)
\prod\limits_{n=0}^{N_-}\dd^2\r_n^-\zeta_+(r_n^-)B'_{N_+,N_-} \label{Xi}
\end{equation}
Indeed, one of the factors $[1-(r^2/4a^2)]^{-1}$ in the area element on
the pseudosphere $\dd S=[1-(r^2/4a^2)]^{-2}\dd^2\r$ has been
incorporated into the definition (\ref{Boltzmann1}) of $B'$, while the
other factor $[1-(r^2/4a^2)]^{-1}$ has been incorporated in the
definition of position-dependent fugacities 
\begin{equation}
\zeta_{\pm}(r)=\frac{\zeta_{\pm}}{1-\frac{r^2}{4a^2}}
\label{fugacities}
\end{equation}
Although the integrals in the grand partition function (\ref{Xi})
diverge (as the separation between a positive particle and a negative
one goes to zero), this grand partition function can be formally
manipulated for providing finite correlation functions. 

It will now be shown that the grand canonical partition function can be
expressed as one determinant of an infinite matrix, continuous in
coordinate space. First, one considers the functional integral
\begin{equation}
Z_0=\int\mathcal{D}\psi\mathcal{D}\bar{\psi}
\int\exp\left[\int\sum_{s,s'=\pm}
\bar{\psi}_s(\r)(M^{-1})_{ss'}(\r,\r')\psi_{s'}(\r')
\dd^2\r\,\dd^2\r'\right]  \label{Z0}
\end{equation}
The fields $\psi$ and $\bar{\psi}$ are two-component Grassmann variables
(anticommuting variables). The components of $\psi$ are called $\psi_+$
and $\psi_-$, and similarly for $\bar{\psi}$. The covariance of the
Gaussian measure in (\ref{Z0}) is the inverse of the kernel $M^{-1}$,
which is chosen such that
\begin{equation}
\langle\bar{\psi}_s(\r)\psi_{s'}(\r')\rangle =M_{ss'}(\r,\r') 
\label{covariance}
\end{equation}
where $\langle\ldots\rangle$ denotes an average taken with the Gaussian
weight of (\ref{Z0}) and the $2\times 2$ matrix $M$ is
\begin{equation}
M(\r,\r')=\left( \begin{array}{cc} M_{++}&M_{+-} \\ M_{-+}&M_{--}
\end{array} \right)=
\left( \begin{array}{cc}
\frac{4a^2}{4a^2-z\bar{z}'}&\frac{2a}{z-z'} \\
\frac{2a}{\bar{z}-\bar{z}'}&\frac{4a^2}{4a^2-\bar{z}z'}
\end{array} \right) \label{M}
\end{equation}
Second, one considers the functional integral
\begin{eqnarray}
Z&=&\int\mathcal{D}\psi\mathcal{D}\bar{\psi}
\int\exp\bigg[\int\sum_{s,s'=\pm}
\bar{\psi}_s(\r)(M^{-1})_{ss'}(\r,\r')\psi_{s'}(\r')\dd^2\r\,\dd^2\r'
\nonumber \\
&+&\int[\zeta_+(r)\bar{\psi}_+(\r)\psi_+(\r)
+\zeta_-(r)\bar{\psi}_-(\r)\psi_-(\r)]\dd^2\r\bigg]  \label{Z}
\end{eqnarray}
and one expands $Z/Z_0$ in powers of $\zeta_+(r)$ and $\zeta_-(r)$ as
\begin{eqnarray}
\frac{Z}{Z_0}&=&
\sum_{N_+=0}^{\infty}\sum_{N_-=0}^{\infty}\frac{1}{N_+!N_-!}\int
\prod\limits_{m=0}^{N_+}\dd^2\r_m^+\zeta_+(r_m^+)
\prod\limits_{n=0}^{N_-}\dd^2\r_n^-\zeta_+(r_n^-)  \nonumber  \\ 
&\times&\langle\bar{\psi}_+(\r_1^+)\psi_+(\r_1^+)\cdots
\bar{\psi}_+(\r_{N_+}^+)\psi_+(\r_{N_+}^+)
\bar{\psi}_-(\r_1^+)\psi_-(\r_1^+)\cdots
\bar{\psi}_-(\r_{N_-}^-)\psi_-(\r_{N_-}^-)\rangle \label{ratio}
\end{eqnarray}
Third, from the Wick theorem for anticommuting variables \cite{Zinn}
and the covariance (\ref{covariance}), it results that the average in 
(\ref{ratio}) is equal to the determinant of the matrix $A_{ij}$ defined 
in (\ref{elements}), i.e to $B'_{N_+,N_-}$ as given by (\ref{Boltzmann2}). 
Therefore (\ref{ratio}) is identical to (\ref{Xi}). The grand partition
function of the Coulomb gas is 
\begin{equation}
\Xi=\frac{Z}{Z_0}    \label{Xi1}
\end{equation}
Finally, $Z_0=\det(M^{-1})$ and $Z=\det(M^{-1}+\zeta)$. In these
determinants of infinite order, the matrix elements of $M$ are labeled
both by the discrete charge indices $s,s'$ and the continuous indices
$\r,\r'$. The infinite diagonal matrix $\zeta$ is defined as
\begin{equation}
\zeta=\left( \begin{array}{cc}
\zeta_+(r)& 0 \\
0         & \zeta_-(r)
\end{array} \right)    \label{zetamatrix}
\end{equation}
Therefore (\ref{Xi1}) does give the grand partition function as the
determinant of an infinite matrix, continuous in coordinate space:
\begin{equation}
\Xi=\det[M(M^{-1}+\zeta)]=\det(1+M\zeta) \label{Xidet}
\end{equation}

For computing the densities and many-body densities, some definitions
are needed. Let us define
\begin{equation} 
\tilde{G}=(1+M\zeta)^{-1}M/(4\pi a)  \label{Gtilde}
\end{equation}
(the factor $(4\pi a)$ is there just for keeping the same notation as in
previous papers). Thus, $\tilde{G}$ is the solution of
$(1+M\zeta)\tilde{G}=M/(4\pi a)$ or, more explicitely, $\tilde{G}$ obeys
the integral equation 
\begin{equation}
\tilde{G}(\r,\r')+\int M(\r,\r'')\zeta(r'')
\tilde{G}(\r'',\r')\dd\r''=\frac{1}{4\pi a}M(\r,\r') \label{inteq}
\end{equation} 
where it should be remembered that $G,M,\zeta$ are $2\times 2$ matrices.
We also define
\begin{equation}
G(\r,\r')=\left(1-\frac{r^2}{4a^2}\right)^{1/2}\tilde{G}(\r,\r')
\left(1-\frac{r'^2}{4a^2}\right)^{1/2} \label{GGtilde}
\end{equation}
On (\ref{M}), one sees the symmetries 
$M_{ss'}(\r,\r')=ss'\bar{M}_{s's}(\r',\r)$. By formally expanding
the definition  $\tilde{G}=(1+M\zeta)^{-1}M/(4\pi a)$ in powers of
$M\zeta$ one finds that $\tilde{G}$ has the same symmetries, which also
hold for $G$:
\begin{equation}
G_{ss'}(\r,\r')=ss'\bar{G}_{s's}(\r',\r) \label{symmetries}
\end{equation}

The density $n_s(\r)$ of particles of sign $s$ is given from the grand
partition function by a functional derivation:
\begin{equation}
n_s(\r)=\left(1-\frac{r^2}{4a^2}\right)^2\zeta_s(r)\frac{\delta\ln\Xi}
{\delta\zeta_s(r)} \label{density1}
\end{equation}
where the factor $[1-(r^2/4a^2)]^2$ insures that $n_s(\r)\dd S$ is the
average number of particles in the area element 
$\dd S=[1-(r^2/4a^2)]^{-2}\dd^2\r$. Since, from (\ref{Xidet}), $\ln\Xi=
\Tr (1+M\zeta)$, (\ref{density1}), (\ref{Gtilde}), and (\ref{GGtilde})
give
\begin{equation}
n_s(r)=4\pi\zeta_s aG_{ss}(\r,\r)  \label{density}
\end{equation}
(actually, for point particles, this density is infinite, but it can 
be made finite by the introduction of a small hard core). The two-body
density Ursell functions are given by
\begin{equation}
U_{ss'}(\r,\r')=\left(1-\frac{r^2}{4a^2}\right)^2
\left(1-\frac{r'^2}{4a^2}\right)^2\zeta_s(r)\zeta_{s'}(r')
\frac{\delta^2\ln\Xi}{\delta\zeta_s(r)\delta\zeta_{s'}(r')}
\label{Ursell1}
\end{equation}
Taking into account the symmetry relations (\ref{symmetries}) gives  
\begin{equation}
U_{ss'}(\r,\r')=-ss'(4\pi\zeta_s a)(4\pi\zeta_{s'} a)|G_{ss'}(\r,\r')|^2
\label{Ursell}
\end{equation}
 From now on, we restrict ourselves to the case of equal fugacities 
$\zeta_+=\zeta_-=\zeta $. In the Poincar\'e disk representation, the
Coulomb fluid fills a disk of radius $r_0$. Thus, $\zeta(r)=0$
when $r>r_0$. The radius $r_0$ is related to the
geodesic radius $R=a\tau_0$ by $r_0=2a\tanh(\tau_0/2)$. Without loss of
generality we can choose the polar angle of $\r'$ as $\varphi'=0$. 

The integral equation (\ref{inteq}) can be tranformed into a differential
one, by the application of the operator $\dbar=\sigma_x\partial_x+
\sigma_y\partial_y$, where $\sigma_x$ and $\sigma_y$ are Pauli matrices:
\begin{equation}
[\dbar+4\pi a\zeta(r)]\tilde{G}(\r,\r')=
\delta_{\mathrm{flat}}^{(2)}(\r-\r')   \label{diffeq} 
\end{equation}
where $\delta_{\mathrm{flat}}^{(2)}$ is the Dirac distribution in the
plane. This differential equation is to be supplemented by the condition
that $\tilde{G}(\r,\r')$ be continuous at the disk boundary $r_0$ and by
the boundary condition, seen on (\ref{inteq}), that when $r=2a$,
$\tilde{G}_{-+}= \e^{\ii\varphi}\tilde{G}_{++}$ (and a similar boundary
relation between $\tilde{G}_{+-}$ and $\tilde{G}_{--}$).

In the case of an infinite system, eq.(\ref{diffeq})
could be solved \cite{beta2}, for $\r'=0$, in terms of hypergeometric
functions. In the present case of a finite disk, i.e. when $r_0<2a$, an
exact explicit solution of (\ref{diffeq}) for an arbitrary fugacity
seems difficult to obtain. Fortunately, here we only need the
large-fugacity limit, in which case there are important simplifications.

\subsection{Large Fugacity}
For a flat system, the Coulomb interaction (\ref{Coulomb}) becomes 
$-\ln(s/2a)$ where $2a$ is an irrelevant length scale which only
contributes an additive constant to the potential. In the flat case
\cite{CJ}, the rescaled fugacity $m=4\pi\zeta a$ (which has the
dimension of an inverse length) was introduced, and the correlation
length was found to be of the order of $m^{-1}$. In the present case 
of a system on a  pseudosphere, it is convenient to keep the same 
definition of $m$. 

On a pseudosphere, in the large fugacity limit $4\pi\zeta a^2=ma\gg 1$,
if we are interested in a solution of (\ref{diffeq}) only in a region of
size $m^{-1}$, the curvature can be neglected and the flat system
solutions can be used, with appropriately rescaled coordinates. In
particular, if both $r$ and $r'$ are sufficiently close to $r_0$, the
variation of $\zeta(r)$ can be neglected: in (\ref{GGtilde}) and 
(\ref{diffeq}), $\zeta(r)$ can be replaced by the constant
$\zeta(r_0)$. Here, we assume the disk to be large, and therefore
$\zeta(r_0)\sim\zeta\e^{\tau_0}/4$. Furthermore (\ref{GGtilde}) becomes
\begin{equation}
\tilde{G}(\r,\r')=\frac{\e^{\tau_0}}{4}G(\r,\r')  \label{GGtilde1}
\end{equation} 
In terms of the rescaled variables
$(\e^{\tau_0}/4)\r=\tbf$ and $(\e^{\tau_0}/4)\r'=\tbf'$,
(\ref{GGtilde1}) and (\ref{diffeq}) do give the flat system 
equation \cite{CJ}
\begin{equation}
[\dbar_{\tbf} +m]G(\r,\r')=\delta_{\mathrm{flat}}^{(2)}(\tbf-\tbf')
\label{Gflat}
\end{equation}

In an infinite system, the $(++)$ and $(-+)$ elements of the solution of
(\ref{Gflat}) would be $G_{++}(\r,\r')=(m/2\pi)K_0(m|\tbf-\tbf'|)$
and $G_{-+}(\r,\r')=(m/2\pi)\e^{\ii\psi}K_1(m|\tbf-\tbf'|)$
where $K_0$ and $K_1$ are modified Bessel functions and $\psi$ is the
argument of $t\e^{\ii\varphi}-t'$. In the present case of a finite
system in a disk, a ``reflected wave'' must be added. 
As a Fourier series in $\varphi$, $G_{++}$ is of the form
\begin{equation}
G_{++}(\r,\r')=\frac{m}{2\pi}\sum_{\ell=-\infty}^{\infty}
\left[I_{\ell}(mt')K_{\ell}(mt)+a_{\ell}I_{\ell}(mt')I_{\ell}(mt)
\right]\e^{\ii\ell\varphi}\;\;\;(t'<t<t_0)  \label{G++}
\end{equation}
where $t_0=(\e^{\tau_0}/4)r_0$. The first term in the sum corresponds to
an expansion\cite{theta} of $K_0(m|\tbf-\tbf'|)$. The second term
corresponds to the ``reflected wave''. The coefficients $a_{\ell}$ are
to be determined by the continuity and boundary conditions. Similarly,
\begin{equation}
G_{-+}(\r,\r')=\frac{m}{2\pi}\sum_{\ell=-\infty}^{\infty}
\left[I_{\ell}(mt')K_{\ell+1}(mt)-a_{\ell}I_{\ell}(mt')I_{\ell+1}(mt)
\right]\e^{\ii(\ell+1)\varphi}\;\;\;(t'<t<t_0)  \label{G-+}
\end{equation}
The corresponding elements of $\tilde{G}$ are given by (\ref{GGtilde1}).
There are similar expansions in the case $t<t'<t_0$.
 
The coefficients $a_{\ell}$ will now be determined. When $t'<t_0<t$, 
(\ref{diffeq}) reduces to $\dbar\tilde{G}(\r,\r')=0$ which means that
$\tilde{G}_{++}$ is an analytic function of $z$ and $\tilde{G}_{-+}$ 
an antianalytic function. Therefore, as a function of
$z=r\e^{\ii\varphi}$, $\tilde{G}_{++}$ is of the form
\begin{equation}
\tilde{G}_{++}=\sum_{\ell=-\infty}^{\infty}
b_{\ell}r^{\ell}\e^{\ii\ell\varphi}\;\;\;(t'<t_0<t) \label{G++out}
\end{equation} 
Taking into account the boundary condition 
$\tilde{G}_{-+}= \e^{\ii\varphi}\tilde{G}_{++}$ at $r=2a$ gives
\begin{equation}
\tilde{G}_{-+}=\sum_{\ell=-\infty}^{\infty}
b_{\ell}(2a)^{2\ell+1}\frac{\e^{\ii(\ell+1)\varphi}}{r^{\ell+1}}
\;\;\;(t'<t_0<t) \label{G-+out}
\end{equation}
For a large disk, $r_0=2a\tanh(\tau_0/2)\sim 2a\exp(-2\e^{-\tau_0})$.
The continuity of $G_{++}$ and $G_{-+}$ at $r=r_0$ determines the
coefficients $a_{\ell}$ and $b_{\ell}$. One finds
\begin{equation}
a_{\ell}=\frac{\exp[-(2\ell+1)2\e^{-\tau_0}]K_{\ell+1}(mt_0)
-K_{\ell}(mt_0)}{\exp[-(2\ell+1)2\e^{-\tau_0}]I_{\ell+1}(mt_0)
+I_{\ell}(mt_0)}  \label{a}
\end{equation}

In the present large-fugacity limit, the Bessel functions in
(\ref{G++}), (\ref{G-+}), and (\ref{a}) can be
replaced by their asymptotic forms $I_{\ell}(x)\sim(2\pi x)^{-1/2}\e^x$ 
and $K_{\ell}(x)\sim(\pi/2x)^{1/2}\e^{-x}$. Then, whatever the relative
order of $t$ and $t'$ might be,
\begin{equation}
G_{++}(\r,\r')\sim\frac{1}{4\pi t_0}\sum_{\ell=-\infty}^{\infty}
\{\e^{-m|t-t'|}-\e^{-m(2t_0-t-t')}
\tanh[(2\ell+1)\e^{-\tau_0}]\}\e^{\ii\ell\varphi}\;\;\;(t,t'<r_0)
\label{G++1}
\end{equation}
and
\begin{equation}
G_{-+}(\r,\r')\sim\frac{1}{4\pi t_0}\sum_{\ell=-\infty}^{\infty}
\{\e^{-m|t-t'|}+\e^{-m(2t_0-t-t')}
\tanh[(2\ell+1)\e^{-\tau_0}]\}\e^{\ii(\ell+1)\varphi}\;\;\;(t,t'<r_0)
\label{G-+1}
\end{equation}
It should be recalled that these expressions are valid only near the
boundary circle.

\subsection{Charge Fluctuations}
For the present model, by symmetry $<Q>=0$ and the variance of the total
charge is 
\begin{equation}
\langle Q^2 \rangle=\int_{r,r'<r_0}\rho^{(2)}(\r,\r')\dd S\,\dd S'
+\int_{r<r_0}n(r)\dd S  \label{chargefluctuation2}
\end{equation}
where $\rho^{(2)}(\r,\r')$ is the two-body charge density, $n(r)$ the
total particle density, and $\dd S$ an area element on the
pseudosphere. In the bulk, perfect screening is 
expected, and furthermore $\rho^{(2)}(\r,\r')$ has a range in the
geodesic distance between $\r$ and $\r'$ of the order of $m^{-1}$ only.
Therefore, the only contributions to (\ref{chargefluctuation2}) come
from $r$ and $r'$ close to $r_0$. When both $r$ and $r'$ are in the
bulk (i.e. smaller enough than $r_0$), $\rho^{(2)}$ becomes a function
$\rho_{\mathrm{bulk}}^{(2)}$ and it is convenient to define a surface
part by $\rho^{(2)}(\r,\r')=\rho_{\mathrm{bulk}}^{(2)}(\r,\r')
+\rho_{\mathrm{surf}}^{(2)}(\r,\r')$. Similarly, the density can be
decomposed as $n(r)=n_{\mathrm{bulk}}+n_{\mathrm{surf}}(r)$. Assuming
that perfect screening occurs in the bulk means 
\begin{equation}
\int\rho_{\mathrm{bulk}}^{(2)}(\r,\r')\dd S+n_{\mathrm{bulk}}=0
\label{screening}
\end{equation}
where the integral extends on the whole pseudosphere. Using
(\ref{screening}) allows to rewrite (\ref{chargefluctuation2}) as
\begin{equation}
\langle Q^2 \rangle=-\int_{r'<r_0<r}\rho_{\mathrm{bulk}}^{(2)}(\r,\r')
\dd S\,\dd S'+\int_{r,r'<r_0}\rho_{\mathrm{surf}}^{(2)}(\r,\r')
+\int_{r<r_0}n_{\mathrm{surf}}(r)\dd S  \label{chargefluctuation3}
\end{equation}
Because of the symmetry between positive and negative particles, 
$\rho^{(2)}(\r,\r')=2[U_{++}(\r,\r')-U_{-+}(\r,\r')]$ with the Ursell
functions given by (\ref{Ursell}) and $n(r)=2n_+(r)$ with $n_+$ given by
(\ref{density}). Thus, (\ref{chargefluctuation2}) becomes
\begin{equation} 
\langle Q^2 \rangle=-2m^2\int_{r,r'<r_0}[|G_{++}(\r,\r')|^2
+|G_{-+}(\r,\r')|^2]\dd S\,\dd S'+2m\int_{r<r_0}G_{++}(\r,\r)\dd S  
\label{chargefluctuation4}
\end{equation}
Using the Fourier series (\ref{G++1}) and (\ref{G-+1}) in
(\ref{chargefluctuation4}), and taking into account that in
$|G_{ss'}|^2$ only the term independent of $\varphi$ survives the
angular integration, gives
\begin{eqnarray}
\langle Q^2 \rangle=&-&\frac{4m^2}{(4\pi t_0)^2}\int_{t,t'<t_0}
\sum_{\ell=-\infty}^{\infty}\{\e^{-2m|t-t'|}+\e^{-2m(2t_0-t-t')}
\tanh^2[(2\ell+1)\e^{-\tau_0}]\}\dd S\,\dd S'  \nonumber \\  
&+&\frac{2m}{4\pi t_0}\int_{t<t_0}\sum_{\ell=-\infty}^{\infty}
\{1-\e^{-2m(t_0-t)}\tanh[(2\ell+1)\e^{-\tau_0}]\}\dd S
\label{chargefluctuation5}
\end{eqnarray}
The first term in each sum corresponds to $\rho_{\mathrm{bulk}}^{(2)}$
and $n_{\mathrm{bulk}}$, respectively, and therefore the second term
corresponds to $\rho_{\mathrm{surf}}^{(2)}$ and $n_{\mathrm{surf}}$,
respectively. Using (\ref{chargefluctuation3}) rather than
(\ref{chargefluctuation2}) gives instead of (\ref{chargefluctuation5})
\begin{eqnarray}
\langle Q^2 \rangle=&-&\frac{4m^2}{(4\pi t_0)^2}
\sum_{\ell=-\infty}^{\infty}\{-\int_{t'<t_0<t}\e^{-2m(t-t')}
\dd S\,\dd S' \nonumber \\
&+&\int_{t,t'<t_0}\e^{-2m(2t_0-t-t')}\tanh^2[(2\ell+1)\e^{-\tau_0}]
\dd S\,\dd S' \} \nonumber \\
&-&\frac{2m}{4\pi t_0}\sum_{\ell=-\infty}^{\infty}\int_{t<t_0}
\e^{-2m(t_0-t)}\tanh[(2\ell+1)\e^{-\tau_0}]\dd S  
\label{chargefluctuation6}
\end{eqnarray}
The integrands are indeed localized near the boundary circle. Using 
$\dd S\sim t_0\,\dd t\,\dd\varphi$ and performing the integrations gives
\begin{equation}
\langle Q^2 \rangle=\frac{1}{4}\sum_{\ell=-\infty}^{\infty}
\{1-\tanh^2[(2\ell+1)\e^{-\tau_0}]\}
-\frac{1}{2}\sum_{\ell=-\infty}^{\infty}\tanh[(2\ell+1)\e^{-\tau_0}]
\label{chargefluctuation7}
\end{equation}
Finally, as $\tau_0$ becomes large, the sums can be expressed as
integrals on the variable $x=(2\ell+1)\e^{-\tau_0}$. Since $\tanh x$ is
an odd function, the second sum can be considered as vanishing (actually, 
there are convergence factors at $\ell\rightarrow\pm\infty$, which
have been omitted when the Bessel functions have been replaced by their
asymptotic forms at fixed $\ell$). One is left with
\begin{equation}
\langle Q^2 \rangle=\frac{\e^{\tau_0}}{8}\int_{-\infty}^{\infty}
(1-\tanh^2 x)\dd x =\frac{\e^{\tau_0}}{4}  \label{chargefluctuation8}
\end{equation}
in agreement with the macroscopic result (\ref{chargevariance1}), since 
here $\beta=2$.

\subsection{Surface Charge Correlations}
The first term in (\ref{G++1}) or (\ref{G-+1}) corresponds to the bulk
contribution $(m/2\pi)K_0(m|\tbf-\tbf'|)$ or $(m/2\pi)\e^{\ii\psi}
K_1(m|\tbf-\tbf'|)$, respectively. The range $m^{-1}$ of these bulk
contributions goes to zero in the large-fugacity limit. Thus, for 
$\tbf\neq\tbf'$, only the second term survives. Let us assume that the
relevant values of $|\varphi|$ are small. Since $\e^{-\tau_0}$ is small
for a large disk, after $(2\ell+1)\e^{-\tau_0}$ has been replaced by
$2\ell\e^{-\tau_0}\sim\epsilon_0\ell$, the sum on $\ell$ can be expressed in
terms of an integral:
\begin{equation}
\sum_{\ell=-\infty}^{\infty}\tanh(\epsilon_0\ell)\e^{\ii\ell\varphi}\sim
\ii\int_{-\infty}^{\infty}\tanh(\epsilon_0\ell)\sin(\ell\varphi)\dd\ell
\label{int}
\end{equation}
Here too, there are convergence factors as
$\ell\rightarrow\pm\infty$, which have been omitted when the Bessel
functions were replaced by their asymptotic forms at fixed $\ell$. These
convergence factors can be taken into account by replacing
$\tanh(\epsilon_0\ell)$ by $\sinh(\epsilon_0\ell)/\cosh(\epsilon\ell)$ 
(with $\epsilon>\epsilon_0$), performing the integral which is a
tabulated one\cite{GR}, and taking the limit
$\epsilon\rightarrow\epsilon_0$ afterwards. The result defines the
integral as   
\begin{equation} 
\ii\int_{-\infty}^{\infty}\tanh(\epsilon_0\ell)\sin(\ell\varphi)\dd\ell=
\frac{\ii\pi}{\epsilon_0\sinh\frac{\pi\varphi}{2\epsilon_0}}
\label{int1}
\end{equation}
The range in $\varphi$ of this function is indeed of the order of
$\epsilon_0$, an a posteriori justification of the above assumption that
$|\varphi|$ is small. Using (\ref{int}) and (\ref{int1}), with
$|\sinh(\pi\varphi/2\epsilon_0)|$ replaced by
$(1/2)\exp\pi|\varphi|/2\epsilon_0)$, in (\ref{G++1}) and (\ref{G-+1})
gives for the two-body charge density near the disk boundary
\begin{equation}  
\rho^{(2)}(\r,\r')=-2m^2[|G_{++}(\r,\r')|^2+|G_{-+}(\r,\r')|^2]=    
-\frac{m^2}{a^2}\e^{-2m(2t_0-t-t')}\exp
\left(-\frac{\e^{\tau_0}\pi|\varphi|}{2}\right) \label{rho2}
\end{equation}
where $\epsilon_0\sim 2\e^{-\tau_0}$ and $t_0\sim a\e^{\tau_0}/2$ have
been used. This two-body charge density is indeed localized near the
disk boundary. The surface charge correlation is defined as 
\begin{equation}
\langle\sigma(\varphi)\sigma(0)\rangle=\int_{-\infty}^{t_0}\dd t
\int_{-\infty}^{t_0}\dd t'\,\rho^{(2)}(\r,\r') \label{correlation2}
\end{equation}
Using (\ref{rho2}) in (\ref{correlation2}) and performing the integrals 
reproduces the macroscopic result (\ref{correlation1}), since here
$\beta=2$ and $<\sigma(\varphi)>=0$.

\section{ONE-COMPONENT PLASMA ON A PSEUDOSPHERE}
The macroscopic results (\ref{chargevariance1}) and (\ref{correlation1})
will now be checked on another solvable model, the one-component plasma.
This is a system of one species of particles, of
charges $+1$, embedded in a uniform background carrying the negative
charge density $-n_b$. At the inverse temperature $\beta=2$, the system
is exactly solvable in a variety of geometries, in particular for a
large disk of radius $R=a\tau_0$ on a pseudosphere \cite{pressures}.
A grand canonical ensemble is used. For the grand partition function to
be convergent, it is necessary to define it with a fixed value\cite{LN}
of the background charge density $-n_b$; the fugacity $\zeta$ controls
the number of particles. Thus, in general, the system is not globally
neutral, except for a particular choice of the fugacity.

In the bulk the properties
of the system are controlled by the background: the particle number
density away from the boundary is $n_b$. However, near the
boundary, the particle density differs from $n_b$, and, since on a
pseudosphere the neighborhood of the boundary has an area 
of the same order of magnitude as the whole area, this neighborhood
gives an important contribution to the total number of particles and
thus to the total charge of the system.

The macroscopic results (\ref{chargevariance1}) and (\ref{correlation1})
are expected to be valid only when the microscopic thickness of the
surface charge density goes to zero. How to reach this regime in the
most general way by varying both parameters $n_b$ and $\zeta$ has not
been clear to us. Here we content ourselves by considering the limit
$\zeta\rightarrow\infty$ for a fixed value of $n_b$. In this limit, the
total charge of the system is expected to become infinite and to be
carried by an infinitely thin surface layer. 

\subsection{Summary of previous results \cite{pressures}}
Again, for two points in the Poincar\'e disk at $\r=(r,\varphi)$ and
$\r'=(r',0)$, one defines an auxiliary quantity $G(\r,\r')$, which now
is just a scalar (instead of a $2\times 2$ matrix). In the case of a
large disk of radius $R=a\tau_0$, $\tau_0\rightarrow\infty$,
\begin{equation}
G(\r,\r')=\zeta\left(\frac{\e^{\tau_0+1}}{4}\right)^{\alpha}
\left(1-\frac{r^2}{4a^2}\right)^{(\alpha +1)/2}
\left(1-\frac{r'^2}{4a^2}\right)^{(\alpha +1)/2}
\sum_{\ell=0}^{\infty}
\left(\frac{rr'}{4a^2}\right)^{\ell}\frac{\e^{\ii\ell\varphi}}
{1+4\pi a^2\zeta\e^{\alpha}\frac{\Gamma(\alpha,x)}{x^{\alpha}}}
\label{GOCP}
\end{equation} 
where $\alpha=4\pi n_b a^2$, $x=4\ell\e^{-\tau_0}$, and
$\Gamma(\alpha,x)$ is the incomplete Gamma function
\begin{equation}
\Gamma(\alpha,x)=\int_x^{\infty} t^{\alpha -1}\e^{-t}\dd t \label{Gamma}
\end{equation}
The particle number density was found to be
\begin{equation}
n(r)=G(\r,\r)  \label{densityOCP}
\end{equation} 
By a similar calculation, one finds for the two-body density Ursell
function
\begin{equation}
U(\r,\r')=-|G(\r,\r')|^2
\label{UrsellOCP}
\end{equation}

In (\ref{GOCP}), $r/(2a)=\tanh(\tau /2)$. Only the case $\tau$ large 
($r$ close to the boundary of the disk) will be needed. Then
$1-[r/(2a)]^2\sim 4\e^{-\tau}$ and
$[r/2a]^{\ell}\sim\exp(-2\ell\e^{-\tau})$. Let us assume that the 
relevant values of $|\varphi|$ are small compared to 1. Then the sum on
$\ell$ can be replaced by an integral on $x=4\ell\e^{-\tau_0}$. This
gives for the density as a function of the distance (in units of $a$)
from the boundary $\lambda=\tau_0-\tau$  
\begin{equation}
\label{eq:resultat-densite-1}
n(\lambda)=G(\r,\r)
=
\zeta \e^{\alpha} \e^{(\alpha+1)\lambda}
\int_0^{\infty}
\frac{\e^{-x\e^{\lambda}}\ \dd x}{\displaystyle
1+ 4\pi a^2 \zeta \e^{\alpha}\frac{\Gamma(\alpha,x)}{x^{\alpha}}}
\end{equation}
Integrating $n(\lambda)$ gives the average number of particles
\begin{equation}
\label{eq:densite-moyenne-1}
\left<N\right>={\cal A}\,
\zeta \e^{\alpha}
\int_0^{\infty}
\frac{\Gamma(\alpha,x)\,\dd x}{\displaystyle
x^{\alpha}+4\pi a^2 \zeta \e^{\alpha}\Gamma(\alpha,x)}
\end{equation}

\subsection{Charge fluctuations}
For the one-component plasma the charge fluctuations are identical (for
particles of charge $+1$) to the particle number fluctuations $\langle
Q^2\rangle^{\mathrm{T}}=\langle N^2\rangle^{\mathrm{T}}$, since the
background charge does not fluctuate. The charge fluctuations can be
obtained either by integrating the correlation function 
(see eq.~(\ref{UrsellOCP})) or by using the thermodynamic relation
\begin{equation}
  \langle N^2\rangle^{\mathrm{T}}
    =\zeta\frac{\partial \langle N\rangle}{\partial \zeta}
\end{equation}
This gives for a large disk
\begin{equation}
  \label{eq:charge-fluctuations-ocp}
  \langle Q^2\rangle^{\mathrm{T}}=
  \frac{\e^{\tau_0}}{4}
  \int_0^{\infty} \frac{ g x^{\alpha}
  \Gamma(\alpha,x)}{\left(x^\alpha+g\Gamma(\alpha,x)\right)^2}\,\dd x  
\end{equation}
where we have defined the dimensionless parameter $g=4\pi a^2 \zeta
\e^{\alpha}$. For any finite value of $\zeta$ and $n_b$ the
integral in the last equation is different from 1, thus the
predictions of macroscopic electrostatics are not satisfied. This is
indeed expected since in general we are out of the validity domain of
macroscopic electrostatics. As explained above, we expect the results
from macroscopic 
electrostatics to be valid only if the thickness of the layer of charge
near the boundary is negligible compared to the macroscopic lengths:
the radius of the disk $R$ and the radius of curvature $a$. For the
two-component plasma the thickness $T$ of this layer is of order of
the inverse of the fugacity $m^{-1}=(4\pi a \zeta)^{-1}$. For the
one-component plasma we shall show that the situation is somehow
different.

Thus, before proceeding to study the charge fluctuations in the 
large-fugacity limit, let us study first how the thickness $T$ of the
charged layer near the boundary depends on $g$ in this limit,
since the situation is not as simple as it is for the two-component
plasma. We will show that indeed $T$ vanishes when $g\to\infty$.

For simplicity let us consider the case when $\alpha=1$. In units of
$a$ the thickness of the charged layer can be defined as the first
moment of the density profile properly normalized
\begin{equation}
T=\frac{\int_0^{\infty} n(\lambda) \lambda \,
\e^{-\lambda}\,\dd\lambda}
{\int_0^{\infty} n(\lambda)\,\e^{-\lambda}\,\dd\lambda}
  =\frac{\int_0^{\infty} n(\lambda) \lambda \,
  \e^{-\lambda}\,\dd\lambda}{n}
\end{equation}
where $n=\left<N\right>/{\cal A}$ is the average particle density.
The $\e^{-\lambda}$ factor comes from the area element 
$\dd S=2\pi a^2 \sinh \tau \dd\tau$ near the boundary. For $\alpha=1$
the density profile~(\ref{eq:resultat-densite-1}) becomes
\begin{equation}
  \label{eq:densite-alpha=1}
  n(\lambda)= n_b\, \e^{2 \lambda}\int_0^{\infty} 
\frac{g x \e^{-x\e^{\lambda}}}{x+g \e^{-x}}\,\dd x
\end{equation}
and the average density is given by
\begin{equation}
  \frac{n}{n_b}= \int_0^{\infty}
  \frac{g\,\e^{-x}}{x+g\e^{-x}}\,\dd x
\end{equation}
Let us define $x_m$ as the principal solution of $g=x_m \e^{x_m}$;
incidentally, the function $x_m(g)$ is the Lambert function, which has
many applications\cite{Lambert}. Now we write
\begin{equation}
  \frac{n}{n_b}= \left(\int_0^{x_m}+\int_{x_m}^{\infty}\right)
  \frac{\dd x}{1+\frac{x}{x_m}\e^{x-x_m}}
\end{equation}
In the first integral ($x<x_m$) the second term in the denominator is
negligible when $g\to\infty$ and then the integrand is 1. After the
change of variable $x\rightarrow x_m+x$, the second integral ($x>x_m$)
is easily shown to have the limit $\ln 2$. This gives in the limit 
$g\to\infty$
\begin{equation}
  \frac{n}{n_b}\sim x_m+\ln 2\sim x_m
\end{equation}
On the other hand, replacing expression~(\ref{eq:densite-alpha=1}) for
the one-body density $n(\lambda)$ into the first moment of the density
and performing the integral over $\lambda$ gives
\begin{equation}
  \int_0^{\infty} n(\lambda) \lambda \, \e^{-\lambda}\,\dd\lambda 
= n_b \int_0^{\infty}
  \frac{g\Gamma(0,x)}{x+ g \e^{-x}}\, \dd x
\end{equation}
Again it is convenient to cut the integral in two intervals for
$x<x_m$ and $x>x_m$. As in the case for $n$ when $g\to\infty$ the
second integral is negligible compared to the first. Then
\begin{eqnarray}
  \int_0^{\infty} n(\lambda) \lambda \, \e^{-\lambda}\,\dd\lambda 
  &\sim& n_b \int_0^{x_m}
  \frac{\e^{x} \Gamma(0,x)}{1+ \frac{x}{x_m} \e^{x-x_m}}\, \dd x
  \nonumber\\
  &\sim& n_b \int_0^{x_m} e^{x} \Gamma(0,x)\, \dd x= \gamma+\e^{x_m}
  \Gamma(0,x_m)+ \ln x_m
\end{eqnarray}
where $\gamma\simeq 0.577$ is the Euler constant. Since, when
$x_m\to\infty$, $\Gamma(0,x_m)\sim \e^{-x_m}/x_m$, the dominant term for
the first moment of the density is the third term in the preceding
equation
\begin{equation}
  \int_0^{\infty} n(\lambda) \lambda \, \e^{-\lambda}\,\dd\lambda \sim 
n_b \ln x_m
\end{equation}
Finally the thickness of the layer of charge near the boundary behaves
as
\begin{equation}
  T \sim \frac{1}{x_m}\ln x_m\to 0
\end{equation}
when $x_m\to \infty$. Remembering that $x_m \e^{x_m}=g=4\pi a^2 \zeta
\e^{\alpha}$ one can notice that the dependence of the thickness on the
fugacity is not trivial. It vanishes when $\zeta\to\infty$ but very
slowly contrarily to the case of the two-component plasma where $T\sim
\zeta^{-1}$.

Now we will proceed to prove that, in the limit $g\gg 1$, the charge
variance~(\ref{eq:charge-fluctuations-ocp}) is equal to the
prediction of macroscopic electrostatics~(\ref{chargevariance1}). To
be as general as possible we consider again any value of $\alpha$. One
can easily prove that the integrand in
Eq.~(\ref{eq:charge-fluctuations-ocp}) is maximum when $x=x_m$ where
$x_m$ is now given by $g=x_m^\alpha/\Gamma(\alpha,x_m)$ for any value
of $\alpha$. Doing the change of variable $x\to x-x_m$ in the
integral~(\ref{eq:charge-fluctuations-ocp}) and replacing $g$ by its
expression in term of $x_m$ gives
\begin{equation}
  \langle Q^2\rangle^{\mathrm{T}}=\frac{\e^{\tau_0}}{4}
  \int_{-x_m}^{\infty}
  \frac{ \left(1+\frac{x}{x_m}\right)^{\alpha}
    \frac{\Gamma(\alpha,x+x_m)}{\Gamma(\alpha,x_m)}}{\left[\left(1+
    \frac{x}{x_m}\right)^{\alpha}+
    \frac{\Gamma(\alpha,x+x_m)}{\Gamma(\alpha,x_m)}\right]^{2}}\, \dd x
\end{equation}
When $g\to\infty$ we have $x_m\to\infty$, $\Gamma(\alpha,x_m)\sim
x_m^{\alpha-1} \e^{-x_m}$ and
\begin{equation}
  \frac{\Gamma(\alpha,x+x_m)}{\Gamma(\alpha,x_m)}\sim
  \left(1+\frac{x}{x_m}\right)^{\alpha-1} \e^{-x}
\end{equation}
Then
\begin{equation}
  \langle Q^2\rangle^{\mathrm{T}}\sim\frac{\e^{\tau_0}}{4}
  \int_{-x_m}^{\infty}
  \frac{ \left(1+\frac{x}{x_m}\right)^{\alpha}
    \left(1+\frac{x}{x_m}\right)^{\alpha-1} \e^{-x}}{\left[\left(1+
      \frac{x}{x_m}\right)^{\alpha}+
      \left(1+\frac{x}{x_m}\right)^{\alpha-1} \e^{-x}
      \right]^{2}}\, \dd x
\end{equation}
We notice that for large values of $|x|$ the integrand vanishes
exponentially as $\e^{-|x|}$. Then, since $x_m\to\infty$, we can
replace the lower limit of the integral by $-\infty$ and neglect
$x/x_m$ in front of $1$. This gives
\begin{equation}
  \langle Q^2\rangle^{\mathrm{T}}\sim\frac{\e^{\tau_0}}{4}
  \int_{-\infty}^{\infty}
  \frac{\e^{-x}\,\dd x}{\left(1+\e^{-x}\right)^2} = \frac{\e^{\tau_0}}{4}
\end{equation}
Since $\beta=2$, this is the expected result~(\ref{chargevariance1})
obtained from macroscopic electrostatics considerations.

\subsection{Surface Charge Correlations}
Under the assumption that the relevant values of $|\varphi|$ are small 
compared to 1, the same manipulations as the ones leading to
(\ref{eq:resultat-densite-1}) give  
\begin{equation}
G(\r,\r')=\zeta \e^{\alpha}\e^{(\alpha +1)\frac{\lambda +\lambda'}{2}}
\int_0^{\infty}
\dd x\frac{\e^{-x\frac{\e^{\lambda}+\e^{\lambda'}}{2}}
\e^{\ii x\frac{\e^{\tau_0}}{4}\varphi}}{1+g\frac{\Gamma(\alpha,x)}
{x^{\alpha}}}
\label{GOCP1}
\end{equation}
where $g=4\pi a^2\zeta\e^{\alpha}$, $\lambda=\tau_0-\tau$, and 
$\lambda'=\tau_0-\tau'$.

Let us consider the case $\varphi >0$. We are interested in the behavior
of (\ref{GOCP1}) as $g\rightarrow\infty$. This behavior will be
shown to be determined by the pole of the integrand closest to the real
axis in the upper half-plane. Let us assume that this pole has a large
real part. Then, at this pole, $\Gamma(\alpha,x)$ behaves as the first
term $x^{\alpha -1}\e^{-x}$ of its asymptotic expansion \cite{GR} and
the denominator in the integrand becomes $1+g\e^{-x}/x$. Let us look for
a zero of this denominator at $x=\ii\pi +x_m$. The equation for $x_m$ is
$(\ii\pi +x_m)\e^{x_m}=g$, which becomes, in the large-$g$ limit, 
$x_m\e^{x_m}=g$. Thus, $x_m$ is large and real and the assumption that
the pole has a large real part is a posteriori verified. The same
reasoning gives other poles at 
$x=(2n+1)\ii\pi +x_m$, $n\in \mathbb{Z}$. The residue of the pole at  
$x=\ii\pi +x_m$ is easily found to have a modulus behaving as 
$\exp[-x_m(\e^{\lambda}+\e^{\lambda'})/2]
\exp[-\pi\e^{\tau_0}\varphi /4]$.

Let $I$ be the integral in (\ref{GOCP1}). $I$ is a part of an
integral in the complex plane, along a contour $C$: $C$ follows the 
positive real axis from 0 to $+\infty$, a large quarter of circle from
$+\infty$ to $+\ii\infty$, and comes back to the origin along the
imaginary axis from $+\ii\infty$ to 0. The contribution from the large
quarter of circle at infinity is easily seen to vanish, and therefore
the contour integral is $I-I'$ with $I'$ the integral with $x=\ii y$ 
pure imaginary varying from 0 to $\ii\infty$:
\begin{equation}
I'=\ii\int_0^\infty\dd y
\frac{\e^{-\ii y\frac{\e^{\lambda}+\e^{\lambda'}}{2}}
\e^{-y\frac{\e^{\tau_0}}{4}\varphi}}
{1+g\frac{\Gamma(\alpha,\ii y)}{\ii y}}
\label{Iprime}
\end{equation}
For large $g$, $I'$ is easily seen to be of order $1/g=\e^{-x_m}/x_m$.

The theorem of residues says that $I=I'+2\pi\ii\,\times$ sum of the 
residues of the poles inside $C$. $I'$ is negligible (by a factor $1/x_m$)
compared to the residue of the pole at $\ii\pi+x_m$. The residues of the
other poles have a factor $\exp[-(2n+1)\pi\e^{\tau_0}\varphi /4],\; n>1$
which makes them also negligible. A similar reasoning holds in the case 
$\varphi <0$, and finally 
\begin{equation}
|G(\r,\r')|\sim 2\pi\zeta\e^{\alpha}\e^{(\alpha +1)
\frac{\lambda +\lambda'}{2}}   
\exp[-x_m\frac{\e^{\lambda}+\e^{\lambda'}}{2}]
\exp(-\pi\frac{\e^{\tau_0}}{4}|\varphi|)
\label{GOCP2}
\end{equation}
This form of $|G|$ a posteriori justifies the assumption that
the relevant values of $\varphi$ are small compared to 1. Furthermore,
in view of the fast decrease of $|G|$ as a function of $\lambda$ or 
$\lambda'$ with a characteristic length $1/x_m$ (compare with the
thickness of $n(r)$ which was found to be $(1/x_m)\ln x_m$), a simpler
form is
\begin{equation}
|G(\r,\r')|\sim 2\pi\zeta\e^{\alpha}   
\exp[-x_m(1+\frac{\lambda}{2}+\frac{\lambda'}{2})]
\exp(-\pi\frac{\e^{\tau_0}}{4}|\varphi|)
\label{GOCP3}
\end{equation}
 
The Ursell function is obtained by using (\ref{GOCP3}) in 
(\ref{UrsellOCP}):
\begin{equation}
U(\r,\r')\sim -(2\pi\zeta\e^{\alpha})^2\exp[-x_m(\lambda +\lambda')]
\exp(-2x_m)\exp(-\pi\frac{\e^{\tau_0}}{2}|\varphi|)
\label{Ursell1OCP}
\end{equation}
The surface charge correlation is defined as
\begin{equation}
\langle\sigma(\varphi)\sigma(0)\rangle^{\mathrm{T}}
=a^2\int_0^{\infty}\dd\lambda\int_0^{\infty}\dd\lambda'\,
U(\r,\r')  \label{correlationOCP}
\end{equation}
Performing the integrations and using 
$4\pi a^2\zeta\e^{\alpha}=x_m\e^{x_m}$ reproduces the macroscopic result
(\ref{correlation1}) at $\beta=2$.

\section{CONCLUSION}
The charge fluctuations for a two-dimensional classical Coulomb fluid
are drastically changed by the introduction of a negative curvature of
space. 

In the case of a flat disk communicating with a reservoir
(grand-canonical ensemble), the total charge $Q$ essentially does not
fluctuate (bringing an additional charged particle from infinity would
cost an infinite energy). In the macroscopic limit, one can define a
surface charge density $\sigma$ (charge per unit length on the boundary
circle). The two-point correlation function of $\sigma$ has an algebraic
only decay (\ref{plane}), behaving as the inverse square distance
between the two points (while the charge correlation function in the
bulk has a faster than algebraic decay).

In the case of a disk on a pseudosphere (an infinite surface of constant
negative curvature), in the macroscopic limit, the total charge $Q$ does
fluctuate with the variance (\ref{chargevariance}). Furthermore the
two-point correlation function of the surface charge density $\sigma$
has a fast (exponential) decay (\ref{correlation1}) as a function of 
the angular distance $|\varphi|$ between the two points.

This change of behavior of the surface charge correlation is related to
the well-known fact that a negative curvature acts as a mass in the
field equations. The curvature replaces the flat logarithmic Coulomb
potential by the potential (\ref{Coulomb}) which has an exponential
decay at large distance $s$. For a flat disk, the algebraic decay of the
two-point surface charge correlation is due to these field lines which 
connect the two points through the vacuum outside the disk. On a
pseudosphere, these field lines outside the disk nevertheless carry an
exponentially decaying interaction. 

For retrieving the macroscopic limit from microscopic models, it is
necessary that the thickness $T$ of the surface charge density be
negligible compared to the macroscopic length scales. On a pseudosphere
with a radius of curvature $a$, in a disk of radius $R$, we have
considered only the case $R\gg a$. The macroscopic behavior is
expected to hold only when $a\gg T$. The two exactly solvable
microscopic models which have been considered do exhibit the expected
macroscopic features when this condition is satisfied. 

\section*{ACKNOWLEDGEMENT}
J.M.Caillol brought to our attention the literature about the Lambert
function. The authors acknowledge support from 
ECOS-Nord/COLCIENCIAS-ICFES-ICETEX action C00P02 of French and Colombian
cooperation.

\end{document}